\documentclass[twocolumn,showpacs,preprintnumbers,amsmath,
               nofootinbib,aps,superscriptaddress]{revtex4}

\usepackage{graphicx}
\usepackage{dcolumn}
\usepackage{bm}

\newcommand*{\no}{\noindent}
\newcommand*{\bea}{\begin{eqnarray}}
\newcommand*{\eea}{\end{eqnarray}}
\newcommand*{\be}{\begin{equation}}
\newcommand*{\ee}{\end{equation}}

\newcommand*{\pref}[1]{(\ref{#1})}

\newcommand*{\mn}{{\mu\nu}}

\newcommand*{\indexsep}{,}
\newcommand*{\tl}{\mathrm{tl}}

\begin{document}

\preprint{}

\title{Two- and three-point Green's functions in two-dimensional Landau-gauge Yang-Mills theory}

\author{Axel Maas}\email{axel.maas@savba.sk}
\affiliation{Department of Complex Physical Systems, Institute of Physics, Slovak Academy of Sciences, D\'{u}bravsk\'{a} cesta 9, SK-845 11 Bratislava, Slovakia}

\date{\today}

\begin{abstract}
The ghost and gluon propagator and the ghost-gluon and three-gluon vertex of two-dimensional SU(2) Yang-Mills theory in (minimal) Landau gauge are studied using lattice gauge theory. It is found that the results are qualitatively similar to the ones in three and four dimensions. The propagators and the Faddeev-Popov operator behave as expected from the Gribov-Zwanziger scenario. In addition, finite volume effects affecting these Green's functions are investigated systematically. The critical infrared exponents of the propagators, as proposed in calculations using stochastic quantization and Dyson-Schwinger equations, are confirmed quantitatively. For this purpose lattices of volume up to (42.7 fm)$^2$ have been used.
\end{abstract}

\maketitle


\section{Introduction}

Two-dimensional Yang-Mills theory turns out to be a very fascinating topic. Quite a number of quantities, e.\ g.\ the string tension \cite{Dosch:1978jt}, can be calculated exactly, although not all quantities are (yet) known analytically. In particular, up to now it was not possible to calculate the Green's functions in Landau gauge. However, exactly these Green's functions may contain interesting information.

The reason for this is confinement. In two-dimensional Yang-Mills theory, confinement in Landau gauge is already manifest in perturbation theory: All elementary fields, the gluons and ghosts, form a BRST quartet, and thus are confined according to the Kugo-Ojima mechanism \cite{Kugo}. This can be extended non-perturbatively, provided that BRST symmetry is unbroken beyond perturbation theory. This makes explicit the absence of propagating degrees of freedom in two-dimensional Yang-Mills theory. But even without propagating degrees of freedom, this permits to investigate the manifestation of the quartet mechanism on the level of the Green's functions.

In addition, the reasoning for the confinement scenario of Gribov and Zwanziger \cite{Gribov,gzwanziger,ZwanzigerFP,Zwanziger} is applicable to two dimensions as well \cite{Zwanziger}. However, this scenario has no direct manifestation on the perturbative level, as in the case of the quartet mechanism. It is only manifest in the infrared properties of correlation functions. In particular, the Gribov-Zwanziger scenario predicts that the Faddeev-Popov operator $M^{ab}$ accumulates near-zero or zero eigenvalues. As a consequence, the ghost propagator $D_G$, being the expectation value of the inverse Faddeev-Popov operator, should be infrared diverging. Detailed calculations using stochastic quantization \cite{Zwanziger} or Dyson-Schwinger equations (DSEs) \cite{Lerche:2002ep,Maas:2004se} lead to a power-law behavior in the far infrared in any dimension from two to four,
\be
D_G(p)\sim_{p\to 0}p^{-2-2\kappa}.\label{dghp}
\ee
\no Furthermore, the gluon propagator is infrared vanishing, and thereby explicitly positivity violating. Its scalar part also behaves like a power-law in the far infrared,
\be
D(p)=\frac{1}{(d-1)}\left(\delta_\mn-\frac{p_\mu p_\nu}{p^2}\right)D_\mn(p)\sim_{p\to 0}p^{-2-2t}\label{dgp},
\ee
\no where $d$ is the space-time dimension. The two exponents are related by the sum rule
\be
t+2\kappa+\frac{4-d}{2}=0\label{dsr}.
\ee
\no Under the assumption of an infrared bare ghost-gluon vertex, two possible values for $\kappa$ are found, 0 and $1/5$ \cite{Zwanziger,Maas:2004se}. If physics is smooth as a function of dimensionality, the non-zero exponent would be expected due to the results obtained in three and in four dimensions \cite{Zwanziger,Lerche:2002ep,Maas:2004se,Zwanziger:2003de}. Note that in calculations using the renormalization group in the case of a bare ghost-gluon vertex the same equations as in DSE calculations are obtained, thus leading to the same results for the infrared exponents in any dimension \cite{Pawlowski:2003hq}.

These two scenarios are two of the most discussed for the confinement mechanism of gluons also in higher dimensions, see e.\ g.\ for four dimensions the reviews \cite{Alkofer:2000wg} and in three dimensions \cite{Zwanziger,Maas:2004se,Cucchieri:2006tf}. A verification of their predictions using lattice gauge theory in higher dimensions has, however, turned out to be very complicated, mainly due to finite volume effects. In three dimensions only a qualitative agreement between the predictions of the Gribov-Zwanziger scenario and functional calculations has been obtained \cite{Cucchieri:2006tf,Cucchieri3d}. In four dimensions, the lattice results are inconclusive (see e.\ g.\ \cite{Cucchieri:2006xi,boucaud,bloch,sternbeck}). Studies using Dyson-Schwinger equations in a finite volume support that these problems are, in fact, finite volume effects, and provide even a quantitative prediction of these in four dimensions \cite{Fischer:2007pf}. The latter are in acceptable agreement with the results obtained in lattice calculations \cite{Fischer:2007pf}.

Here, for two dimensions, the accessible lattices permit a quantitative test of the predictions. It will be shown that the predictions, assumptions, and actually the value of $\kappa=1/5$, of the Gribov-Zwanziger scenario are found in lattice calculations, and hence there is very strong evidence for the Gribov-Zwanziger scenario to be at work. In fact, it is possible to quantify the finite volume effects.

Hence, in the following a quantitative confirmation of the predictions of the Gribov-Zwanziger scenario using lattice gauge theory for two-dimensional SU(2) Yang-Mills theory in (minimal) Landau gauge will be given.

Of course, with such results, one question immediately arises when comparing the two-dimensional results to those in higher dimensions: Why do they agree qualitatively on the level of two- and three-point Green's functions in the infrared? This points to a structural origin of both, the Gribov-Zwanziger and the Kugo-Ojima scenario, provided both are, in fact, correct. It is particularly tempting to then investigate the relation of both scenarios in two dimensions. Also how the relation of two-dimensional Yang-Mills theory to topological field theory  \cite{Birmingham:1991ty} comes then into play is immediately on one's mind. These, and similar questions arise when contemplating the results, and indicate that there are many interesting opportunities still present in the study of two-dimensional Yang-Mills theory. These are highly interesting questions, and must be investigated in the future.

Within this work, however, as a first step just the results from the lattice calculations will be collected. The two-point functions, and as associated quantities the Faddeev-Popov operator and the running coupling, will be investigated in section \ref{sprop}. The three-point functions will afterwards be discussed in section \ref{svertx}. A short summary of the results will be given in section \ref{ssum}. The technicalities of the lattice simulations can be found in appendix \ref{agen}. Lattice artifacts other than finite volume effects will be discussed in appendix \ref{aarti}. That the suppression of color indices in equations \pref{dghp} and \pref{dgp} is justified will be shown in appendix \ref{atensor}.

\section{Two-point functions}\label{sprop}

The definition and determination of the two-point functions on the lattice, and the associated quantities, have been repeatedly discussed in the literature (see, e.\ g., \cite{ZwanzigerFP,Cucchieri:2006tf,Cucchieri:2006xi,boucaud,bloch,sternbeck}). Here, the methods described in \cite{Cucchieri:2006tf} are employed. Furthermore, the appearance of $\beta$-factors to obtain the correct scaling has been discussed there, also in case of the three-point functions. Hence, this will not be repeated here. To assign units to the quantities, the exactly calculable string tension \cite{Dosch:1978jt} has been assigned the conventional value (440 MeV)$^2$, as in higher dimensions.

\subsection{Gluon propagator}

\begin{figure}
\includegraphics[width=\linewidth]{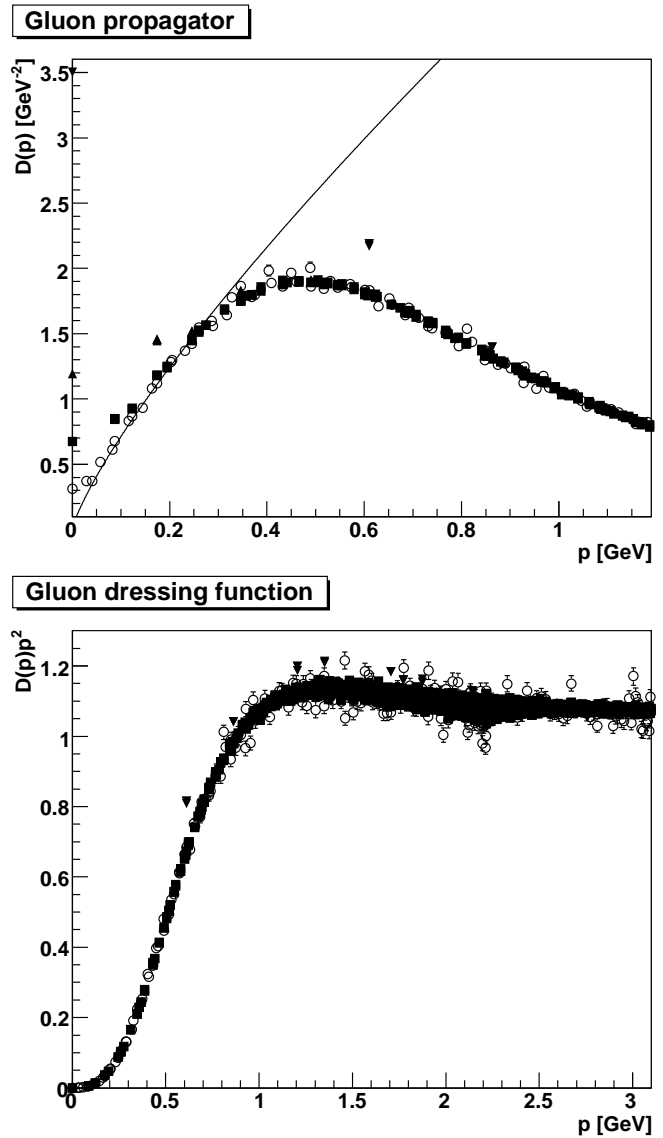}
\caption{\label{fgp}The top panel shows the gluon propagator at small momenta for various volumes. The lower panel shows the gluon dressing function over the whole accessible momentum range. Open circles correspond to a volume of (42.7 fm)$^2$, full squares to (14.2 fm)$^2$, full triangles to (7.11 fm)$^2$, and upside-down full triangles to (2.02 fm)$^2$. The solid line in the top panel is the function $4.5p^{4/5}$.}
\end{figure}

\begin{figure}
\includegraphics[width=\linewidth]{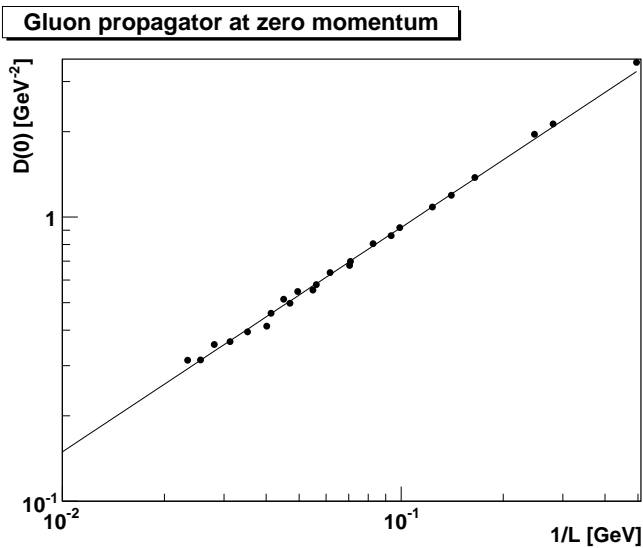}
\caption{\label{fd0}The zero-momentum value $D(0)$ of the gluon propagator as a function of inverse edge length. The straight line is the power-law fit $5.67L^{-0.79}$ to the 20 points at the smallest volumes.}
\end{figure}

The gluon propagator is the most readily accessible two-point correlation function. The results for the propagator $D(p)$, \pref{dgp}, and its associated dressing function $p^2D(p)$ are shown in figure \ref{fgp}. A strongly infrared suppressed gluon propagator is clearly visible. At the same time, the infrared suppression increases with increasing physical volume. In particular, while on a volume of (2.02 fm)$^2$ the propagator appears to be infrared diverging, a clear maximum appears already at a volume only a factor (2-3)$^2$ larger. Only the point at the lowest non-vanishing momentum and the point at zero are not consistent with an infrared vanishing gluon propagator. This is, however, expected \cite{Fischer:2007pf}. The scaling of $D(0)$ with volume, shown in figure \ref{fd0}, makes it very likely that in the infinite volume limit the gluon propagator vanishes at zero momentum, as it vanishes like a power-law with inverse volume. In fact, the exponent $0.79$ of the determined power-law is in very good agreement with the expectation \cite{Fischer:2007pf} that it should coincide with the exponent of the gluon propagator $t=4/5$ of equation \pref{dgp}.

Furthermore, even the gluon dressing function does not exhibit any qualitative difference to three dimensions. In particular it also exhibits a (shallow) maximum. As the propagator becomes ultraviolet constant, as a consequence of asymptotic freedom, there is no intrinsic necessity for such a maximum, as in four dimensions. Its presence in this theory without propagating degrees of freedom is hence slightly surprising. However, in the context of a DSE treatment, it is natural to expect such a maximum due to the different signs of ghost and gluon self-energy contributions \cite{Maas:2004se}.

\begin{figure}
\includegraphics[width=\linewidth]{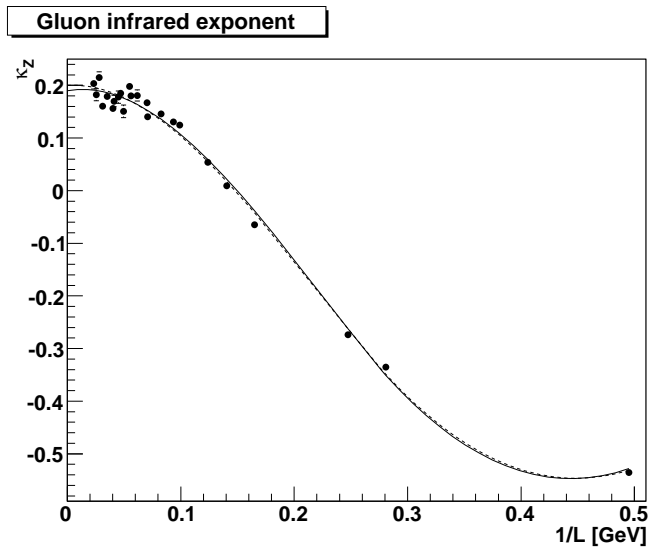}
\caption{\label{fgpex}The measured infrared exponent $\kappa_Z$ obtained from the gluon propagator. Two fits are given. The dashed line corresponds to a fit of type \pref{gpfit} which is forced to go to the predicted value $\kappa=1/5$ at $1/L=0$, while the one given by the solid line is not forced to do so. The fit parameters can be found in table \ref{tgpfit}.}
\end{figure}

The most interesting quantity is the far infrared behavior. It is clearly visible that the gluon propagator is strongly infrared suppressed. The deviation at the very lowest momenta points, however, shows a more massive behavior, as expected from DSE-studies in finite volumes \cite{Fischer:2007pf}. However, the mass decreases rapidly with volume, as discussed above, and a massive behavior is seen only in a momentum window which rapidly decreases with increasing volume. More interestingly, it is expected that in the regime\footnote{The characteristic scale $\Lambda_{\mathrm{QCD}}$ is in two dimensions of course proportional to the coupling constant $g$.}
\be
\frac{2\pi}{aN}\ll p\ll \Lambda_{\mathrm{QCD}}\label{doa}
\ee
\no the continuum behavior should prevail. In particular, the gluon propagator should decrease even in a finite volume in this domain like the power-law \pref{dgp} \cite{Fischer:2007pf}. Using the sum rule \pref{dsr}, the exponent of the propagator itself should be $4\kappa$. Such a power-law is shown in the top panel in figure \ref{fgp}, and agrees well with the data inside the domain \pref{doa}.

To investigate this quantitatively, the effective exponent $\kappa_Z$ was determined. This was done by discarding the two lowest non-vanishing momentum points. Then the next five highest points in momentum were used to fit a power-law. To obtain errors, the steepest and shallowest curve consistent with a 1$\sigma$-confidence interval was determined as well. That this is likely too optimistic is shown by the scattering of the results below. If more than one momentum representation for a given momentum existed, the results were averaged over the various representations, as the violation of rotational symmetry is a minor effect that far in the infrared, see appendix \ref{aarti}. The results are shown in figure \ref{fgpex}. While there are still significant fluctuations at large volumina, the measured exponents tend towards the continuum value.

\begin{table}[h]
\caption{\label{tgpfit}Fit parameters of formula \pref{gpfit}. Fit $1$ corresponds to one with fixed $a=\kappa=1/5$, fit $2$ to one where $a$ was fitted as well.}
\begin{ruledtabular}
\vspace{1mm}
\begin{tabular}{|c|c|c|c|c|}
Fit & $a$ & $b$ [fm] & $c$ [fm$^2$] & $d$ [fm$^3$] \cr
\hline
1 & 1/5 & 0.130 & -12.9 & 19.5 \cr
\hline
2 & 0.190 & 0.358 & -14.0 & 20.9 \cr
\hline
\end{tabular}
\end{ruledtabular}
\end{table}

The volume-dependence of the measured exponents can be fitted by the formula\footnote{The cubic term is necessary to include all volumes.}
\be
\kappa_Z^f=a+\frac{b}{L}+\frac{c}{L^2}+\frac{d}{L^3}.\label{gpfit}
\ee
\no Two fits have been done. In one case, $a$ was fitted as well, while in the second case $a$ was set to the continuum value $\kappa=1/5$. However, even with $a$ free, the result is in reasonable agreement with $1/5$. In particular, the results are not consistent with an infrared finite gluon propagator, which would be expected if $\kappa=0$, the second solution found in \cite{Zwanziger,Maas:2004se}. The individual fit parameters are given in table \ref{tgpfit}.

Hence, the gluon propagator behaves quantitatively exactly as predicted in the Gribov-Zwanziger scenario, when finite volume effects are taken properly into account.

\subsection{Ghost propagator}

\begin{figure}
\includegraphics[width=\linewidth]{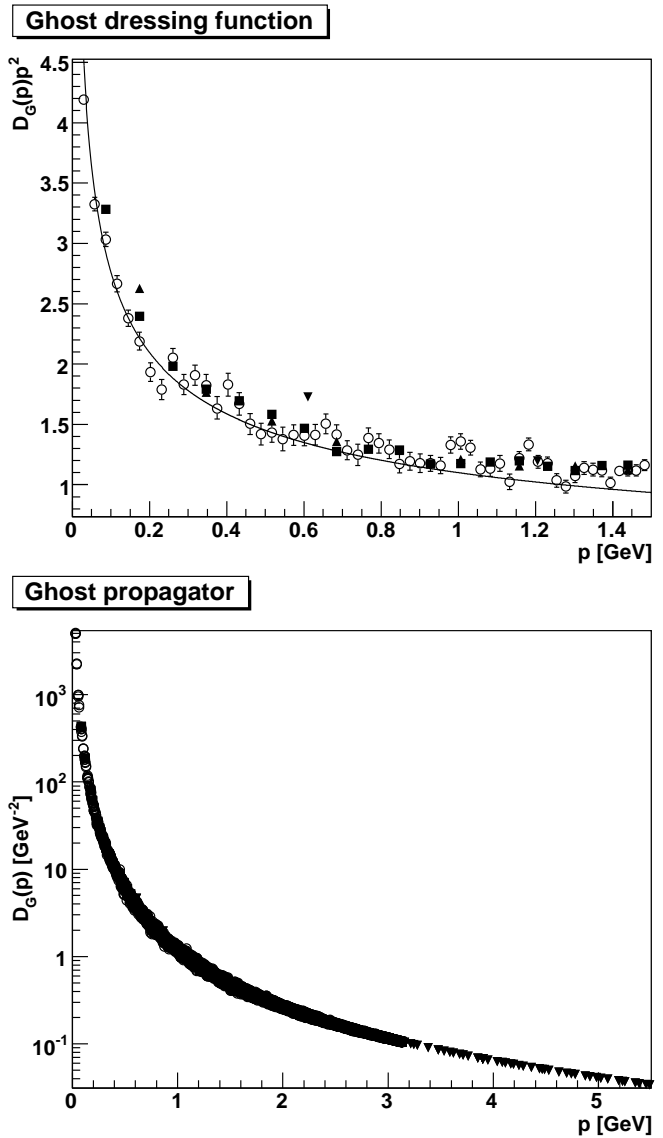}
\caption{\label{fghp}The top panel shows the ghost dressing function at small momenta for various volumes. The lower panel shows the ghost propagator over the whole accessible momentum range. Open circles correspond to a volume of (42.7 fm)$^2$, full squares to (14.2 fm)$^2$, full triangles to (7.11 fm)$^2$, and upside-down full triangles to (2.02 fm)$^2$. The solid line is the function $1.1p^{-2/5}$.}
\end{figure}

The ghost propagator has been determined along the same line as in higher dimensions \cite{ZwanzigerFP,Cucchieri:2006tf}. However, more interesting than the propagator itself is the dressing function $p^2 D_G(p)$. The propagator and the dressing function are shown for different volumes in figure \ref{fghp}. It is clearly visible that the dressing function is infrared diverging. This already indicates that of the two possible exponents $\kappa=0$ and $\kappa=1/5$ found \cite{Zwanziger,Maas:2004se} only the latter one, if one at all, is realized.

Compared to the case of the gluon propagator, finite volume effects are hardly visible to the eye. It seems that the propagator actually becomes less infrared diverging with volume. From the quantitative evaluation below, this is found to be not the case. What seems to be the case is that the domain of closest approach to the origin is affected by finite volume effects. Its modification leads to the various changes in the infrared in a non-trivial manner. If this is the case, the finite-volume effects would be very hard to compare between lattice calculations and functional calculations, as they would be dominated by mid-momentum effects, which in functional methods are usually most strongly affected by truncations \cite{Maas:2004se,Alkofer:2000wg}. This would, on the other hand, explain why in four dimensions the finite volume effects in the ghost propagator have indeed been found to be at least to some extent different in lattice and in Dyson-Schwinger calculations \cite{Fischer:2007pf}. In addition, Gribov-Singer effects \cite{Gribov,Singer:dk}, which according to the Gribov-Zwanziger scenario are irrelevant in the infinite-volume limit \cite{Zwanziger:2003cf}, may still be relevant even at volumes as large as those used here. This has not yet been investigated in two dimensions in Landau gauge.

Even with the available volumes the effect is small. A power-law with exponent $\kappa=1/5$ already describes the data quite well in the infrared, as shown in the top panel of figure \ref{fghp}. Therefore, a more quantitative investigation of the infrared behavior is required.

\begin{figure}
\includegraphics[width=\linewidth]{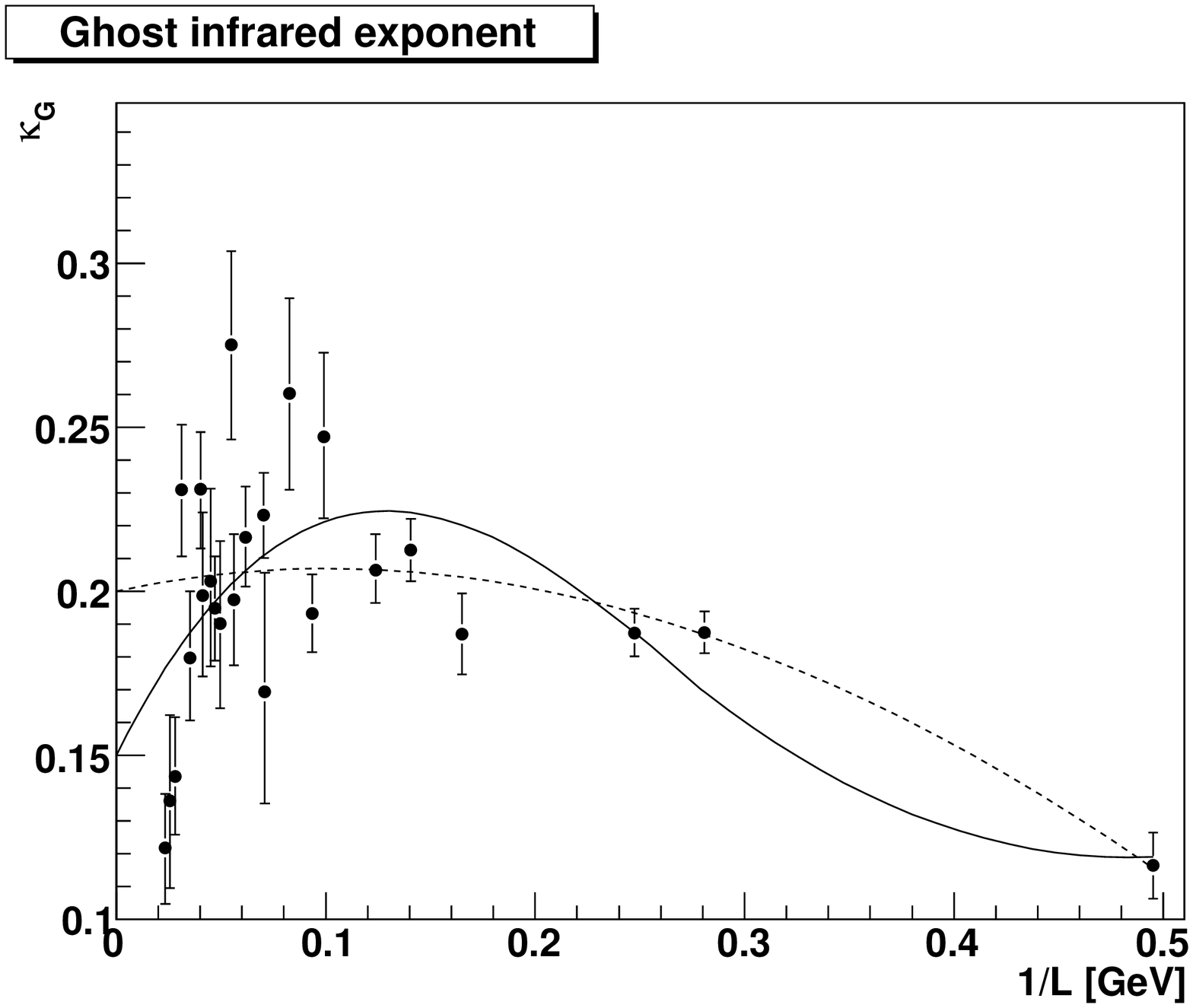}
\caption{\label{fghpex}The measured infrared exponent $\kappa$ obtained from the ghost propagator. Two fits of type \pref{gpfit} are given. The dashed line corresponds to a fit which is forced to go to the predicted value at $1/L=0$, while the one given by the solid line is not forced to do so. The fit parameters can be found in table \ref{tghpfit}.}
\end{figure}

\begin{table}[h]
\caption{\label{tghpfit}Fit parameters for the ghost effective exponent $\kappa_G$ using formula \pref{gpfit}. Fit $1$ corresponds to one with fixed $a=\kappa=1/5$, fit $2$ to one where $a$ was fitted as well.}
\begin{ruledtabular}
\vspace{1mm}
\begin{tabular}{|c|c|c|c|c|}
Fit & $a$ & $b$ [fm] & $c$ [fm$^2$] & $d$ [fm$^3$] \cr
\hline
1 & 1/5 & 0.139 & -0.711 & 0.173 \cr
\hline
2 & 0.150 & 1.28 & -6.41 & 7.47 \cr
\hline
\end{tabular}
\end{ruledtabular}
\end{table}

This is done by extracting the effective infrared ghost exponent $\kappa_G$ in the same way as in the case of the gluon propagator. The results for $\kappa_G$ are shown in figure \ref{fghpex}. While statistical errors are larger than in the case of the gluon propagator, it is visible that all results cluster around the predicted continuum value of $\kappa=1/5$ at large volumes. This is also seen from a fit of the type \pref{gpfit}. The corresponding fit parameters can be found in table \ref{tghpfit}. Due to statistical uncertainties it is not as clean as for the gluon. However, it is visible that the exponent does not vary strongly with volume. In fact, the effective ghost exponent seems only to change by about a third when changing the volume by almost two orders of magnitude. Finally, even with the limited fit accuracy it is not unreasonable that the results are, in fact, consistent with the prediction of $\kappa=1/5$.

\begin{figure}
\includegraphics[width=\linewidth]{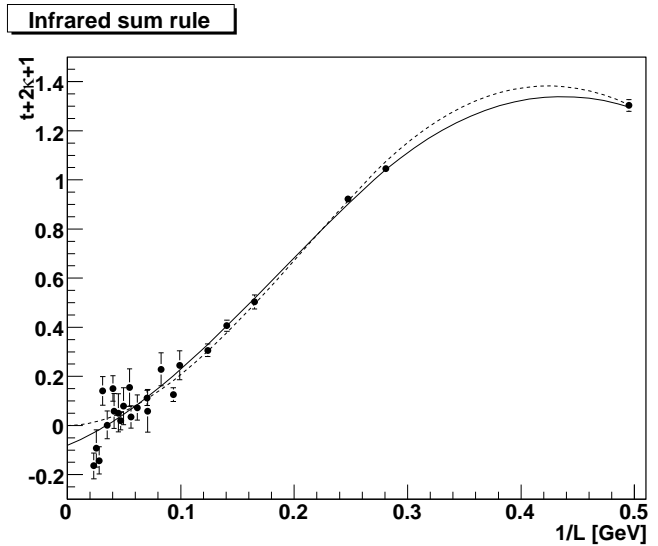}
\caption{\label{fsumrule}Test of the sum rule $t+2\kappa+1=0$, using the effective ghost exponent $\kappa_G$, shown in figure \ref{fghpex}, and the effective gluon exponent $t_Z=-(1+2\kappa_Z)$, shown in figure \ref{fgpex}. Two fits of type \pref{gpfit} are given. The dashed line corresponds to a fit which is forced to go to the predicted value at $1/L=0$, while the one given by the solid line is not forced to do so. The fit parameters are given in table \ref{tsumrulefit}.}
\end{figure}

Another possibility to check the continuum results is to test the predicted sum rule \pref{dsr}. This is done by using the effective measured exponents $\kappa_Z$ and $\kappa_G$ in figure \ref{fsumrule}. Again, a fit of type \pref{gpfit} has been performed. The corresponding fit parameters are given in table \ref{tsumrulefit}. As already anticipated from the individual results, the sum rule becomes better and better satisfied when approaching the continuum limit. Hence, it seems very likely that the shiny relation \pref{dsr} is, in fact, recovered in the continuum limit.

\begin{table}[h]
\caption{\label{tsumrulefit}Fit parameters for a formula of type \pref{gpfit} for the sum rule. Fit $1$ corresponds to one with fixed $a=0$, fit $2$ to one where $a$ was fitted as well.}
\begin{ruledtabular}
\vspace{1mm}
\begin{tabular}{|c|c|c|c|c|}
Fit & $a$ & $b$ [fm] & $c$ [fm$^2$] & $d$ [fm$^3$] \cr
\hline
1 & 0 & 0.0192 & 24.4 & -38.6 \cr
\hline
2 & -0.0806 & 1.85 & 15.2 & -26.9 \cr
\hline
\end{tabular}
\end{ruledtabular}
\end{table}

\begin{figure}
\includegraphics[width=\linewidth]{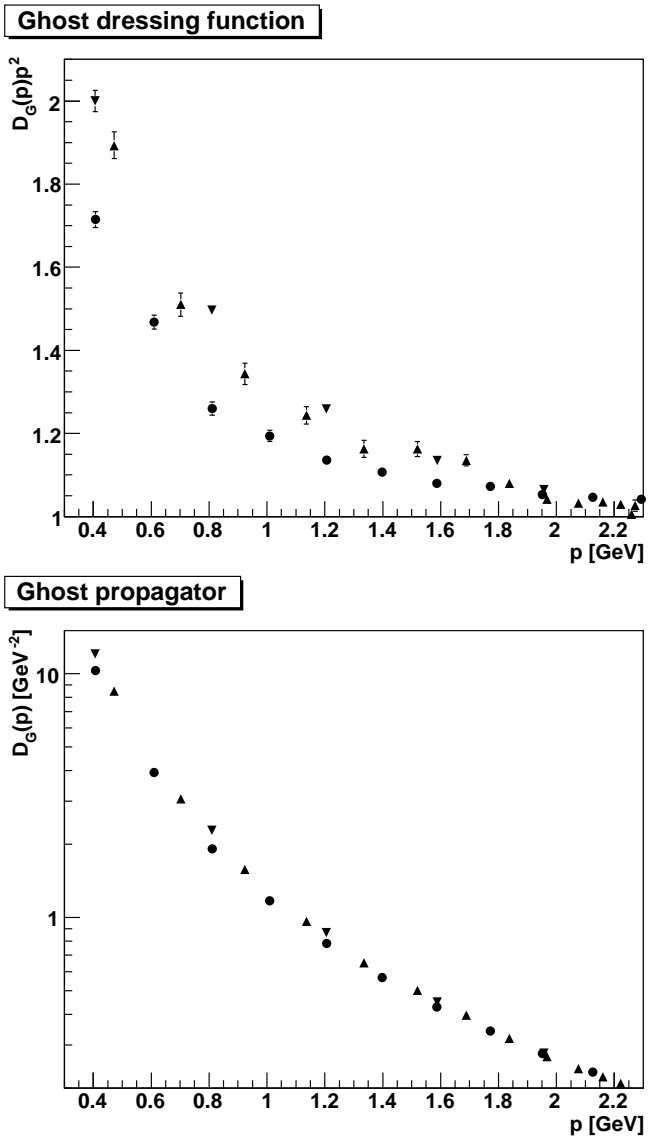}
\caption{\label{fghp-nd}Comparison of the ghost propagator in different dimensions. Circles are two dimensions, triangles are three dimensions, and upside-down triangles are four dimensions. The lattice volumes are (6.06 fm)$^2$, (5.20 fm)$^3$ \cite{Cucchieri:2006tf}, and (5.28 fm)$^4$ \cite{ftv}, at $\beta=30$, $\beta=4.2$, and $\beta=2.3$, respectively.}
\end{figure}

One of the particularly interesting results so far is that the ghost exponent is only very weakly dependent on the volume, compared to the one of the gluon. This is in marked contrast to the case in four-dimensional DSEs in a finite volume \cite{Fischer:2007pf}. Furthermore, all attempts to extract a ghost exponent in lattice calculations in higher dimensions also yield a rather small, more or less volume-independent exponent \cite{sternbeck}. It is thus interesting to compare the ghost propagator in various dimensions at roughly the same volume. This is done in figure \ref{fghp-nd}. Only the momentum regime is shown which is accessible by all of the lattices used. Furthermore, the propagators have been normalized so that they coincide at a momentum of 2 GeV. For the momenta itself, the string tension was set to the same value for all three different dimensionalities.

Aside from the question to which extent such a comparison is justified, the results behave as predicted: The ghost propagator becomes more divergent with increasing dimension. Also, it is in agreement with the predictions \cite{Zwanziger,Lerche:2002ep,Maas:2004se,Alkofer:2000wg} that the difference is more pronounced from two to three dimensions than from three to four dimensions: $\kappa$ changes from $1/5$ to $\approx 0.39$ or $1/2$ from two to three dimensions. The four-dimensional exponent of $\kappa\approx 0.59$ is, on the other hand, rather close to the one in three dimensions. This qualitative behavior, with all its caveats, is another indication for the correctness of the Gribov-Zwanziger scenario and the quantitative predictions.

Furthermore, the result in two dimensions is in fact confirming quantitatively the Gribov-Zwanziger scenario in two dimensions. However, the very slow change in the effective exponent over orders of magnitude in volume is indicative of what challenges have to be met in higher dimensions to see the asymptotic ghost exponent.

Finally, the exact value of the exponent obtained in DSE calculations depends on the projection of the tensor equation for the ghost \cite{Maas:2004se,Alkofer:2000wg}. The value of 1/5 is obtained only in the case of a transverse projection \cite{Maas:2004se}. This in turn implies automatically a certain structure of the longitudinal (w.\ r.\ t.\ to the gluon momentum) tensor structure of the ghost-gluon vertex, such that it leads for arbitrary projections to the infrared exponent 1/5. This then makes the determination of this tensor structure an almost trivial exercise in the infrared limit. Furthermore, this precisely prescribes how the Slavnov-Taylor identity for the gluon propagator, and hence its transversality, is recovered in the far infrared.

\subsection{Running coupling}\label{src}

Although it is possibly a questionable concept in two-dimensional Yang-Mills theory, it is possible to formally define a running coupling. Analogous to higher dimensions \cite{Alkofer:2000wg,vonSmekal}, the quantity\footnote{To improve the statistical behavior, the ghost dressing function has been evaluated on a plane-wave source instead of a point source, as in case of the propagator alone \cite{Cucchieri:2006tf}. Hence only the same volumes are accessible for the coupling constant as for the ghost-gluon vertex below, where this is also necessary.} $\alpha(p)=p^6D(p)D_G(p)^2$ is then proportional to the coupling constant. In particular, as a consequence of the sum rule, the quantity $\alpha(p)$ should behave in the infrared as $p^2$. Hence $\alpha/p^2$ should be constant.

\begin{figure}
\includegraphics[width=\linewidth]{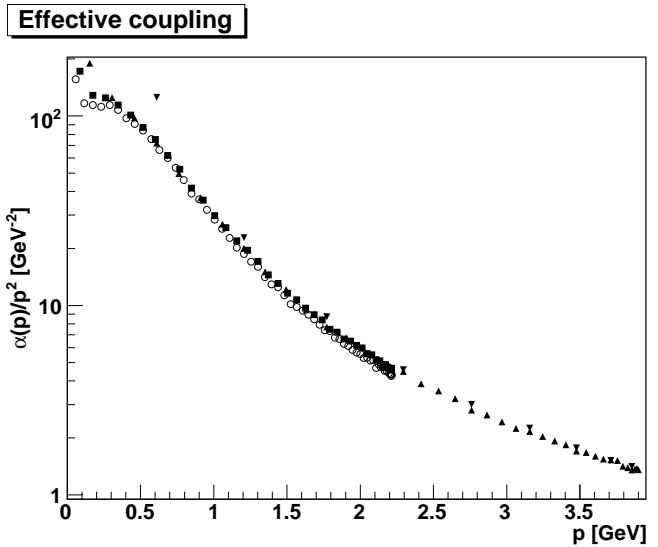}
\caption{\label{falpha}The effective running coupling divided by $p^2$ for various volumes. Open circles correspond to a volume of (21.3 fm)$^2$, full squares to (14.2 fm)$^2$, full triangles to (8.08 fm)$^2$, and upside-down full triangles to (2.02 fm)$^2$.
}
\end{figure}

From the results on the sum-rule, given in figure \ref{fsumrule}, it is already clear that an infrared fixed point will hardly be seen. However, the results, shown in figure \ref{falpha}, exhibit such a fixed point at the largest volumes, provided the lowest point at non-vanishing momentum is discarded\footnote{For the coupling constant only edge momenta have been used, in contrast to the propagators where also other momenta have been included. Dismissing here only the lowest non-vanishing momenta is thus equivalent to dismissing the two lowest non-vanishing momenta in case of the propagators.}. Note that the finite volume effects seem to make the running coupling diverging instead of vanishing, as in higher dimensions \cite{sternbeck,Fischer:2007pf}.

Thus at sufficiently large volumes, and taking finite volume effects into account, it is in fact possible to observe a fixed point in the coupling in lattice gauge theory.

Note that there is a small, systematic overall factor between the coupling obtained in the different volumes shown in figure \ref{falpha}. This effect is not visible in the propagators themselves, but is increased here by taking effectively the third power of the propagators. As this effect occurs at all momenta, it is likely not simply a finite volume effect. However, this can still be an ${\cal O}(a)$-effect which is caused, e.\ g. among other effects, by the fact that tadpole corrections, which give overall-factors to the propagators, have been neglected here \cite{bloch,Lepage:1992xa}.

\subsection{Faddeev-Popov operator}

A last element in the analysis of the two-point correlation functions are the properties of the Faddeev-Popov operator, central to the Gribov-Zwanziger scenario \cite{Gribov,gzwanziger,ZwanzigerFP}. The results on the ghost propagator, which is the expectation value of the inverse Faddeev-Popov operator, already indicate the existence of an enhancement of its eigenspectrum near zero eigenvalue. This enhancement is the hallmark of the Gribov-Zwanziger scenario. However, it is interesting to see the quantitative behavior of the eigenspectrum. Hence the spectral properties of the Faddeev-Popov operator have been determined as well, using the technique described in \cite{Cucchieri:2006tf}.

\begin{figure}
\includegraphics[width=\linewidth]{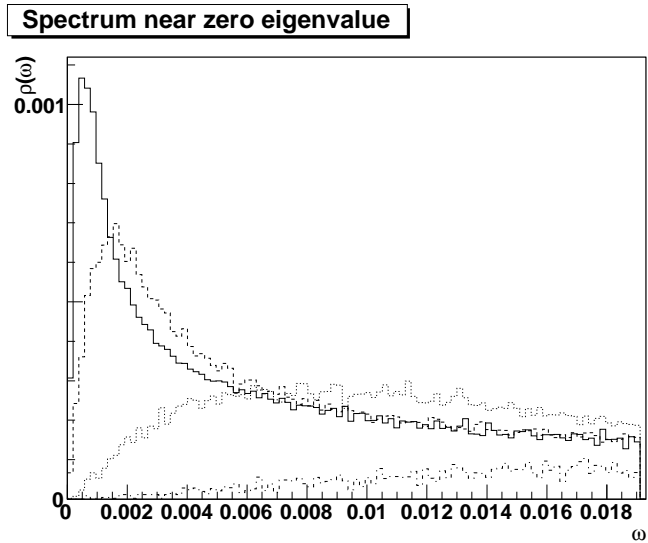}
\caption{\label{ffpspec}The near-zero part of the eigenvalue spectrum of the Faddeev-Popov operator for volumes (2.02 fm)$^2$ (dashed-dotted line), (7.11 fm)$^2$ (dotted line), (14.2 fm)$^2$ (dashed line), and (24.9 fm)$^2$ (solid line). 1164228, 2261493, 3517400, and 2614098 eigenvalues have been enclosed in the full spectrum, respectively.}
\end{figure}

\begin{figure}
\includegraphics[width=\linewidth]{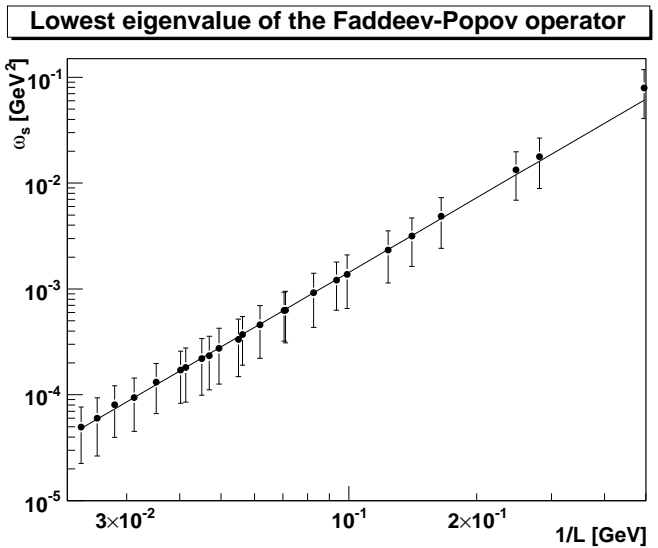}
\caption{\label{flowev}The volume-dependence of the lowest eigenvalue of the Faddeev-Popov operator. The solid line is the function  $0.314L^{-2.34}$.}
\end{figure}

The near-zero part of the eigenspectrum is shown for various volumes in figure \ref{ffpspec}. The volume scaling of the lowest eigenvalue is shown in figure \ref{flowev}. It is clearly visible that with increasing volume more and more eigenvalues are found near zero. This is the near-zero eigenvalue enhancement, as predicted in the Gribov-Zwanziger scenario\footnote{Note that the decrease towards larger eigenvalues seen in figure \ref{ffpspec} is likely an artifact of the method to determine the eigenvalues \cite{Cucchieri:2006tf}. Furthermore, all eigenvalues are only found with multiplicity 1.}. In addition, the lowest eigenvalue vanishes in the infinite-volume limit, and in fact vanishes faster than the lowest eigenvalue of the Laplacian. This is a property which has also been observed in higher dimensions \cite{Cucchieri:2006tf,ftv}. It has been argued that such a larger rate may be necessary for the ghost propagator to develop an infrared divergence \cite{Cucchieri:2006hi}. It is therefore another direct evidence for the validity of the Gribov-Zwanziger scenario. This vanishing of the lowest eigenvalue is in fact necessary for the Gribov-Zwanziger mechanism to work: For infinite volume, an average configuration should be arbitrarily close or on the common boundary of the fundamental modular region and the Gribov horizon, where by definition the determinant of the Faddeev-Popov operator vanishes, and thus must have at least one vanishing eigenvalue \cite{ZwanzigerFP}.

\section{Three-point functions}\label{svertx}

Investigating the vertices in two dimensions is a very interesting task. On the one hand, the vertices do not lend themselves easily to evaluation, since as three-point functions they are much more strongly affected by statistical fluctuations than two-point functions. Hence their investigation has so far been limited to rather small volumes in four \cite{Maas:2006qw,Cucchieri:2004sq} and even in three dimensions \cite{Cucchieri:2006tf,Maas:2006qw}. On the other hand, the vertices describe interaction effects, and it is not a-priori clear how they should behave in a theory without propagating degrees of freedom. In particular, the possibility that the three-gluon vertex, or at least some of its tensor structures, could change sign is a very interesting observation in higher dimensions \cite{Cucchieri:2006tf,Maas:2006qw}. Whether this is also the case in two dimensions, especially in large volumes, is thus also a question of interest.

One drawback of investigating vertices in two dimensions on a square lattice is the impossibility to construct a momentum configuration such that all three momenta are equal. This equal momentum configuration is the one usually preferred in functional studies of the vertices \cite{Fischer:2006vf}, as it permits to have only one external scale. However, in higher dimensions it was found that the results do not change qualitatively when instead two of the momenta are taken to be orthogonal \cite{Cucchieri:2006tf,Maas:2006qw}. This configuration can be realized in two dimensions, and will thus be employed here.

In general, vertices have a significant amount of tensor structures. To obtain a more simple function to measure the interaction represented by a vertex, the quantity
\be
G=\frac{\Gamma^{\tl\indexsep abc}G^{abc}}{\Gamma^{\tl\indexsep abc}D^{ad}D^{be}D^{cf}\Gamma^{\tl\indexsep def}}\label{vdef}.
\ee
\no will be evaluated instead. Here the indices $a, \ldots, f$ are generic multi-indices, encompassing field-type, Lorentz and color indices. Also, $D^{ab}$ are the propagators of the fields, $G^{abc}$ represent the full Green's functions and $\Gamma^{\tl\indexsep abc}$ are the corresponding tree-level vertices. This quantity is defined such that it becomes equal to one if the full and the tree-level vertex coincide. For a more detailed discussion of this quantity and its properties, see \cite{Cucchieri:2006tf}.

\begin{figure*}[th]
\begin{minipage}[c]{0.5\linewidth}
\includegraphics[width=\linewidth]{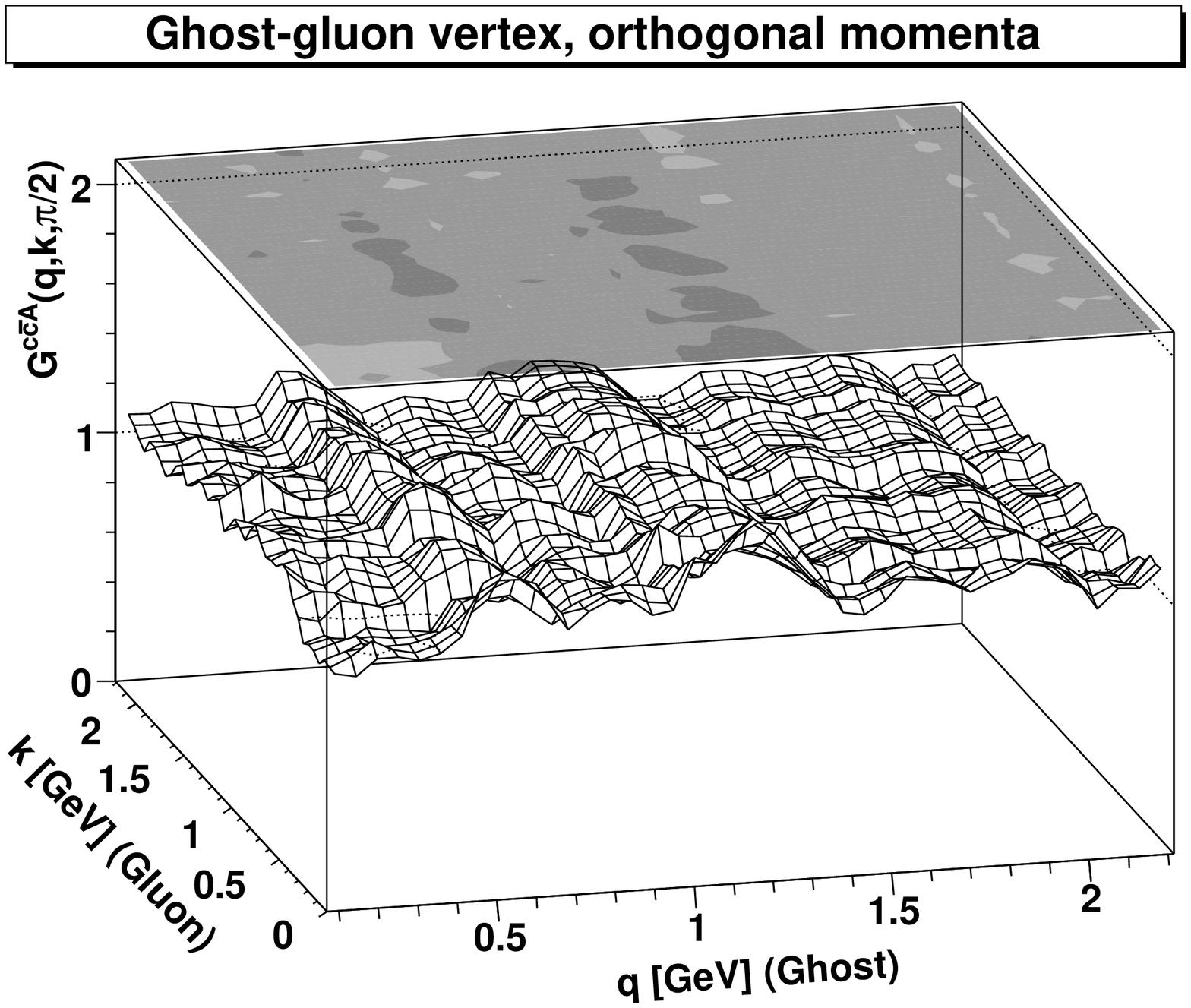}
\end{minipage}%
\begin{minipage}[c]{0.5\linewidth}
\caption{\label{fggva} The ghost-gluon vertex for orthogonal momenta. The top left panel shows the vertex for all possible orthogonal momentum configurations for a volume of (21.3 fm)$^2$, with errors suppressed. The ripple structure is an artifact of the method \cite{Cucchieri:2006tf}, and vanishes with increasing statistics. The bottom left and right panel show the vertex in two specific momentum configurations. In one case the gluon momentum vanishes (left panel), and in the other the gluon and ghost momenta are of equal magnitude (right panel). In this case, various physical volumes are compared. Open circles correspond to a volume of (21.3 fm)$^2$, full squares to (14.2 fm)$^2$, full triangles to (8.08 fm)$^2$, and upside-down full triangles to (2.02 fm)$^2$.}
\end{minipage}\\
\begin{minipage}[c]{\linewidth}
\includegraphics[width=\linewidth]{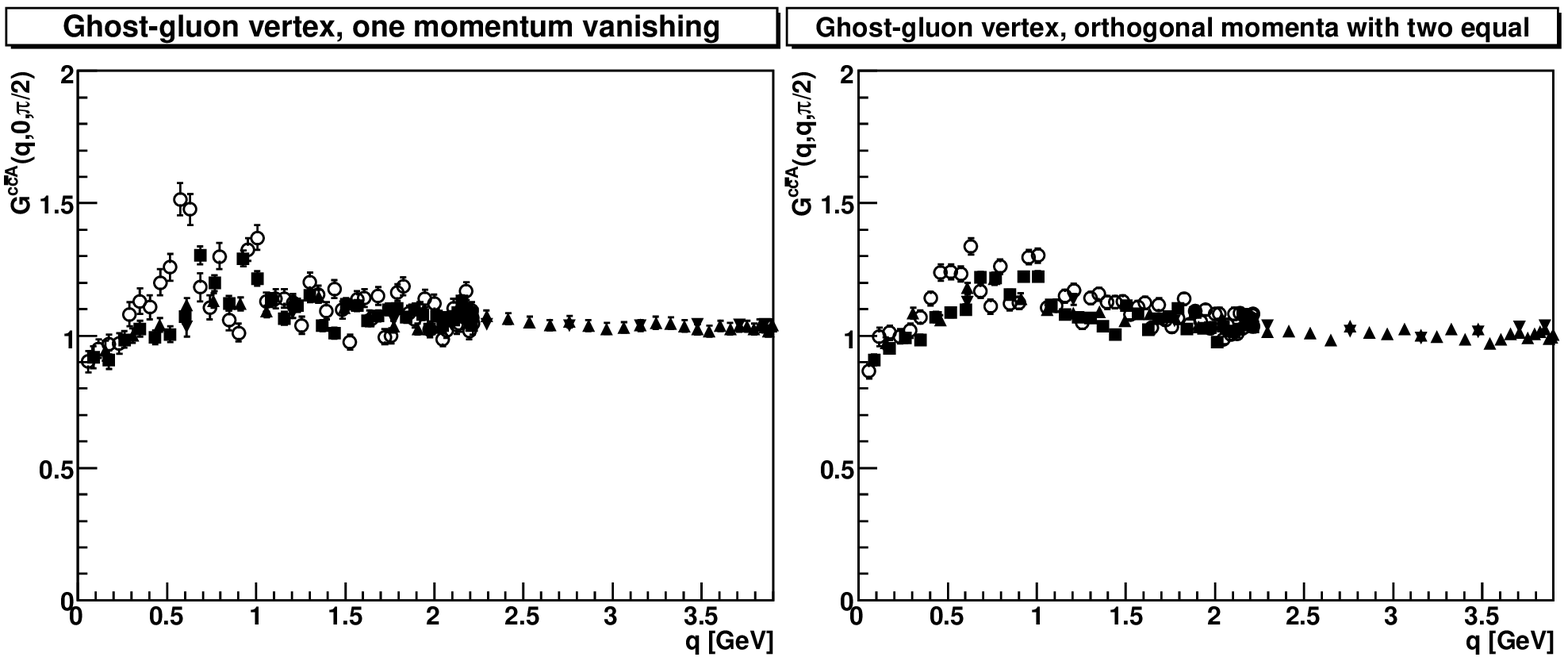}
\end{minipage}
\end{figure*}

There are two vertices in Landau-gauge Yang-Mills theory. The first is the ghost-gluon vertex, which is shown for four different volumes in figure \ref{fggva}. In this case in fact the vertex is shown, as only one tensor structure, the tree-level one, survives non-amputation \cite{Cucchieri:2006tf}.

As in the higher-dimensional cases \cite{Cucchieri:2006tf,Maas:2006qw,Cucchieri:2004sq}, it exhibits an essentially constant behavior, except for a possible small structure below roughly 1 GeV in ghost momentum. This structure is a maximum, with a drop towards smaller momenta below the tree-level value. Furthermore, the value at small ghost momenta and finite gluon momenta is below 1, but finite. If a modification away from a constant infrared behavior of this vertex should exist, it must set in with an extremely small effective exponent to not be visible on these volumes.

These results are all in qualitative agreement with the ghost-gluon vertex in higher dimensions  \cite{Cucchieri:2006tf,Maas:2006qw,Cucchieri:2004sq}. In particular, the results confirm the truncation scheme in the far infrared used in two dimensions in stochastic quantization and DSE calculations \cite{Zwanziger,Lerche:2002ep,Maas:2004se}. In that case an infrared finite ghost-gluon vertex was assumed, delivering the critical infrared exponent $\kappa=1/5$, which in fact was observed in the previous section. This once more nicely confirms the Gribov-Zwanziger scenario, which leads directly to this type of approximation. Furthermore, in four dimensions the infrared critical exponent of the ghost-gluon vertex is fixed, once the exponents of the propagators are known \cite{Fischer:2006vf}. This can be extended to two dimensions and yields in fact an infrared constant ghost-gluon vertex \cite{fipriv}. This is a very stringent test of the scenario. The results found here in lattice calculations once more pass this test. Or, more aptly put, the test passes the results.

\begin{figure*}
\includegraphics[width=\linewidth]{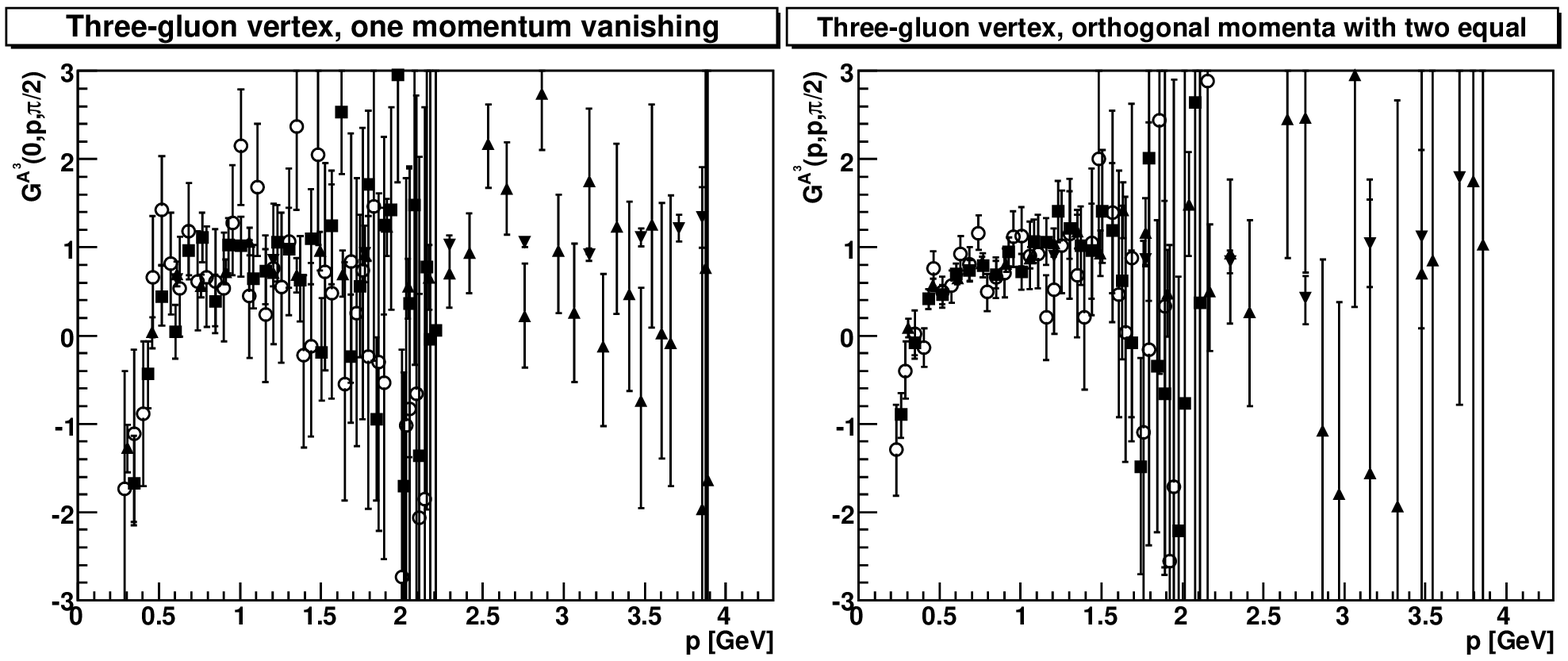}\\
\includegraphics[width=0.5\linewidth]{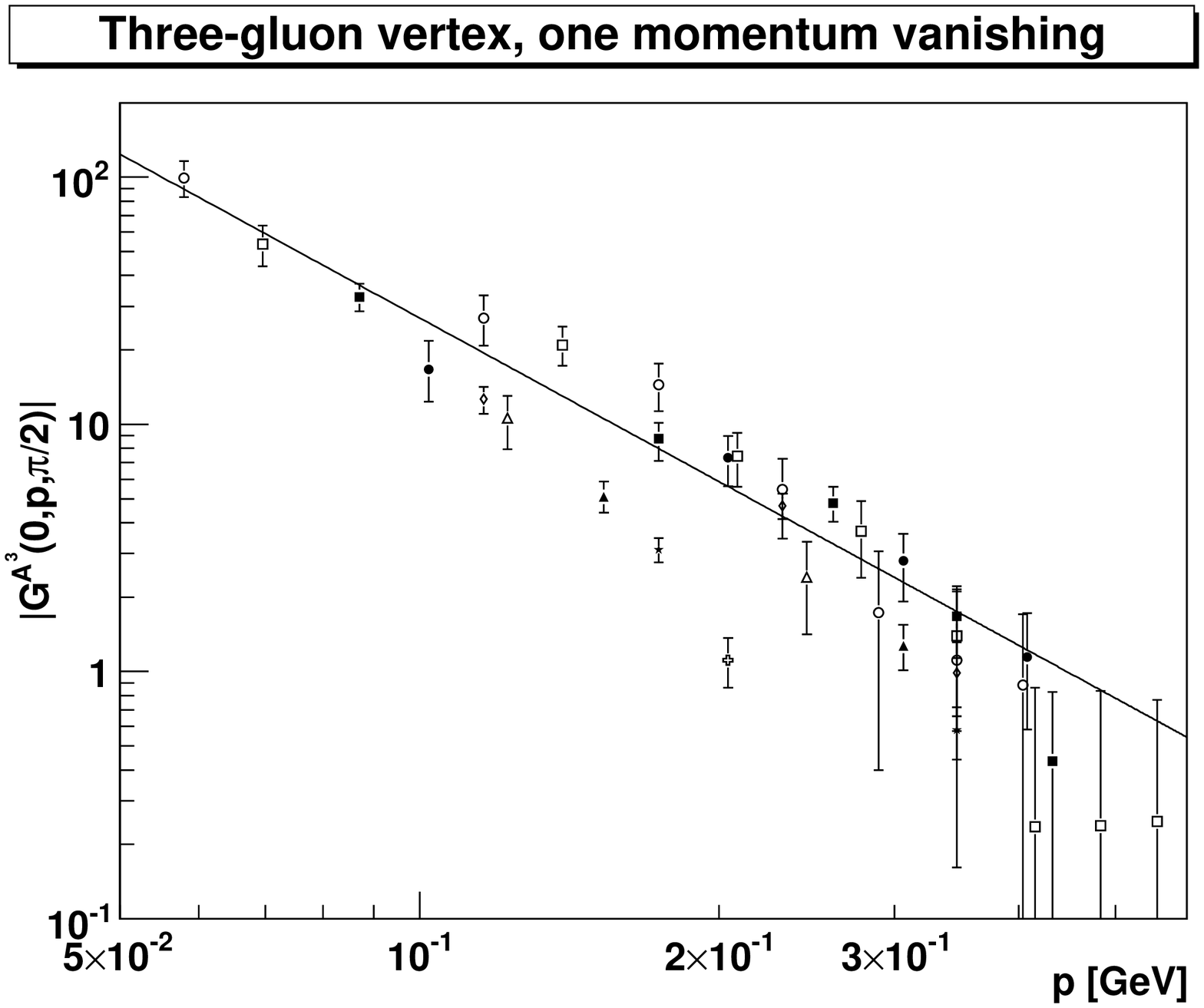}\includegraphics[width=0.5\linewidth]{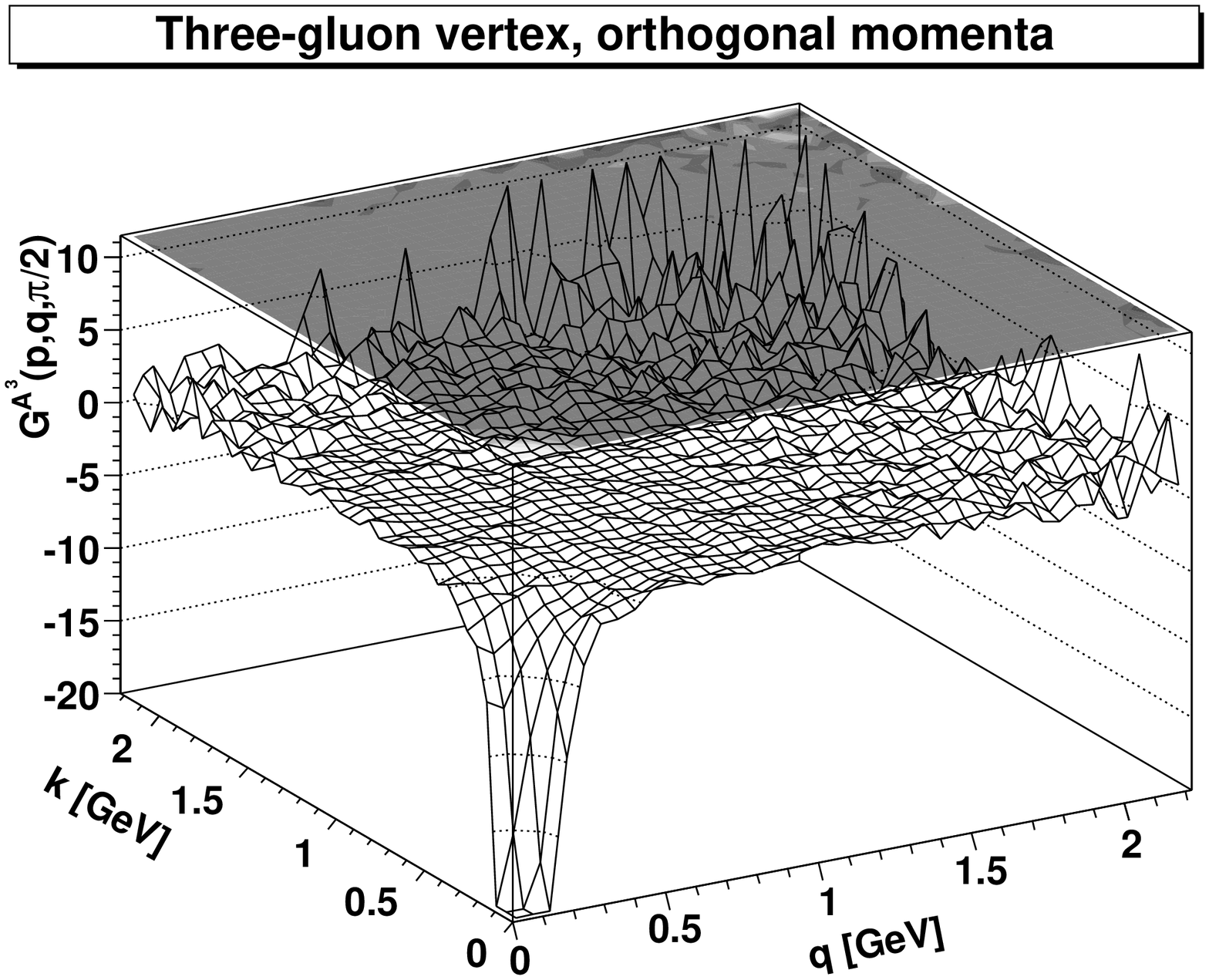}
\caption{\label{fg3va} The three-gluon vertex for orthogonal momenta. The top left and right panel show the vertex in two specific momentum configurations. In one case one of the gluon momenta vanishes (left panel), and in the other two of the gluon momenta are of equal magnitude (right panel). The bottom left panel shows a magnification of the low-momentum regime for one momentum vanishing. In this case the absolute value of $G^{A^3}$ is displayed. Various physical volumes are compared. Open circles correspond to a volume of (21.3 fm)$^2$, full squares to (14.2 fm)$^2$, full triangles to (8.08 fm)$^2$, and upside-down full triangles to (2.02 fm)$^2$. Finally, in the bottom right panel $G^{A^3}$ is shown for the complete orthogonal momentum configuration plane in case of the largest volume (21.3 fm)$^2$. In case of the bottom left panel, results from all available volumes up to lattices of size 120$^2$ are shown, see appendix \ref{agen}. In addition to the previously used symbols, the remaining symbols correspond to (3.56 fm)$^2$ (pluses), (4.04 fm)$^2$ (open stars), (6.06 fm)$^2$ (open crosses), (7.11 fm)$^2$ (full stars), (10.1 fm)$^2$ (open triangles), (10.7 fm)$^2$ (diamonds), (12.1 fm)$^2$ (full circles), and (17.8 fm)$^2$ (open squares). The line is the function $-0.17p^{-2.2}$.}
\end{figure*}

The three-gluon vertex is much more troublesome to calculate due to strong statistical fluctuations, in particular at large lattice (not physical) momenta. These are, in fact, even more pronounced than in higher dimensions, as was already observed when going from three to four dimensions \cite{Maas:2006qw}. Thus the uncertainty connected with this vertex is quite large. Nonetheless, the results shown in figure \ref{fg3va} are quite spectacular. At a point of about 300-400 MeV, corresponding roughly to the position where the plateau in the coupling constant develops or where the gluon propagator starts to bend over, the quantity $G^{A^3}$ changes sign. Thereafter, it diverges, likely like a power-law, as can be seen from the bottom-left panel in figure \ref{fg3va}. Precisely such a divergence is expected in higher dimensions \cite{Fischer:2006vf}. This also compares very well to lattice results in higher dimensions, which found the onset of such a negative divergence in three dimensions \cite{Cucchieri:2006tf}, and at least an infrared suppression in four dimensions \cite{Maas:2006qw}. Note, however that due to the contraction \pref{vdef} not necessarily one particular tensor structure of the vertex changes sign. It is as well possible that two tensor structures have opposite sign throughout, but differ in magnitude, and one is dominant in the infrared, while the other dominates in the ultraviolet.

The infrared divergence of the three-gluon vertex when one momentum vanishes is roughly in agreement with a power-law with exponent $-2.2$ for the single external scale, as can be seen in the bottom-left panel of figure \ref{fg3va}. Although this is not the momentum configuration used in DSE calculations \cite{Fischer:2006vf}, there is again just one external scale. It could be expected that the infrared behavior is the same, if there is just one scale left. In that case, this exponent of $-2.2$ is actually the one expected in DSE calculations \cite{fipriv}. This statement applies as well to the infrared constancy of the ghost-gluon vertex.

Taking this reasoning seriously would imply that all two- and three-point functions exhibit exactly and quantitatively the infrared exponents predicted in DSE calculations, and are in agreement with the Gribov-Zwanziger scenario. Therefore, this work here would represent the first quantitative confirmation of these two frameworks using lattice gauge theory.

It is of course tempting to also investigate higher $n$-point functions. Unfortunately, this is currently out of reach in the present approach. The reason is that only non-amputated, full Green's functions can be directly obtained with the methods used here. Therefore, it would be necessary to first subtract the not-connected part of the amplitude, and then amputate the Green's functions. In general, the not-connected and the connected amplitude have the same infrared behavior, at least in four dimensions, if the predictions \cite{Fischer:2006vf} are correct. Therefore, it would be necessary to disentangle the sum of two functions, which both have the same leading infrared behavior. As the statistical fluctuations become larger when increasing the number of external legs, the required statistics become impractical at the current time. Hence it would be necessary to reduce these fluctuations. It is possible that e.\ g.\ including only results for the same sign of the Polyakov loop\footnote{At finite volume, the value of the Polyakov loop is non-zero for each individual configuration.} would be helpful, as by this statistical fluctuations, at least in case of the gluon propagator, are reduced \cite{Cucchieri:1997dx}. This has to be investigated further.

\section{Summary}\label{ssum}

The volumes accessible in two-dimensional Yang-Mills theory permitted here to obtain the two-point and three-point functions on very large lattices, up to (42.7 fm)$^2$ and (21.3 fm)$^2$, respectively. In particular, it was possible to obtain quantitative results on the infrared behavior with a precision which is unprecedented in the lattice investigations of these quantities.

These results demonstrated that the gluon propagator is infrared vanishing, the ghost propagator is infrared diverging, and the 'effective coupling constant' also has the expected qualitative infrared behavior. Moreover, it was possible to make these statements quantitative. Including the effects of finite volume, it was possible to determine the infinite-volume limit of the characteristic infrared exponents for the two-point functions, and demonstrate the validity of the sum-rule \pref{dsr}. In fact, the value $\kappa=1/5$ found coincides with one of the two possible values expected from stochastic quantization and Dyson-Schwinger equations for an 'on-shell', i.\ e.\ transverse, gluon. Furthermore, the infrared behavior of the vertices permit to close the system self-consistently in the context of such equations. In particular, the ghost-gluon vertex is infrared constant.

These results confirm the Gribov-Zwanziger scenario in two dimensions. Without any dynamic, i.\ e.\ propagating, degrees of freedom, all the infrared behavior is still qualitative the same as in higher dimensions. This implies that these effects in fact stem from the gauge-fixing procedure, in essentially the way predicted by the Gribov-Zwanziger scenario.

It will, of course, take some time before it is possible to repeat the same in higher dimensions. One of the quantitative reasons is that the critical exponent in the gluon observables decreases with increasing dimension \cite{Zwanziger,Maas:2004se}. Hence the effects observed here will only be observable for larger volumes in higher dimensions. Nonetheless, the results are also in excellent qualitative agreement with the predictions of DSE calculations for the finite volume behavior of the propagators in four dimensions \cite{Fischer:2007pf}. Finally, the comparison of the ghost propagator for different dimensions yields the pattern expected from the Gribov-Zwanziger scenario.

However, these results should also be taken with care, as two-dimensional Yang-Mills theory is different from its higher-dimensional versions. And although there is little evidence to the contrary, no rigorous implication exists that the effects seen here translate themselves into higher dimensions without changes. Hence a satisfactory state of affairs in higher dimensions has to await equivalent investigations in higher dimensions. Until then, these results here are another piece of the puzzle, which seem to indicate that the Gribov-Zwanziger scenario in Landau gauge is valid also in higher dimensions.

These results are, beyond these questions, also interesting on their own. It is very tempting to investigate how these results relate to the host of exact results available in two-dimensional Yang-Mills theory, what is the connection to the topological aspects of the theory, and, last but not least, how and if an equivalence between the Gribov-Zwanziger and the Kugo-Ojima confinement scenario exists, at least in two dimensions.

\acknowledgments
 
The author is grateful to Attilio Cucchieri and Tereza Mendes for many helpful and interesting discussions. Furthermore he thanks all those (in particular, Markus Huber) who always ask about two dimensions. This work was supported by the DFG under grant MA 3935/1-2 and in part by the Slovak Grant Agency for Science, Project VEGA No.\ 2/6068/2006. The ROOT framework \cite{Brun:1997pa} has been used in this project.

\appendix

\section{Generation of configurations}\label{agen}

\begin{table*}[ht]
\caption{\label{conf}Data of the configurations considered in the numerical simulations. The values for $a$ are 1.108 GeV$^{-1}$ for $\beta=10$ and 1.951 GeV$^{-1}$ for $\beta=30$ \cite{Dosch:1978jt}. The momenta $p_0$, $p_i$, and $p_f$ denote the lowest non-vanishing momentum and the beginning and the end of the fit interval used in the determination of the effective exponents in section \ref{sprop}, respectively. Note that for $N\ge 140$ not as many momentum configurations for the gluon propagator were available as for $N\le 120$. Ghost configurations are the ones used to determine the ghost propagator, the properties of the Faddeev-Popov operator, the ghost-gluon vertex, and the running coupling. Gluon configurations are the ones used to determine the gluon propagator and the three-gluon vertex. As the autocorrelation time for the plaquette is less than one hybrid overrelaxation (HOR) sweep, all sweeps (after thermalization) have been used for the plaquette measurement, given the number of plaquette configurations in the table. Note that all ghost configurations are also included in the gluon configurations, the sets are not independent. In case of $N\ge 140$, only the propagators have been determined. Hence the number of both configurations coincide. The quantity $<P>/<P_\infty>$ gives the ratio of the expectation value of the plaquette over the analytical infinite volume limit. The error is determined according to \cite{Cucchieri:2006tf}. Finally, $p$ is the tuning parameter for the stochastic overrelaxation algorithm used for gauge-fixing \cite{Cucchieri:1995pn}, and which has been obtained by linear self-adjustment \cite{Cucchieri:2006tf}. Note that this quantity is not very precisely determined, and should be used rather as an indication of the correct order. Sweeps is the number of HOR sweeps between two consecutive measurements \cite{Cucchieri:2006tf}.}
\begin{ruledtabular}
\vspace{1mm}
\begin{tabular}{|c|c|c|c|c|c|c|c|c|c|c|c|}
$V$ [fm$^2$] & $N=\sqrt{V/a^2}$ & $\beta$ & $p_0$ [MeV] & $p_i$ [MeV] & $p_f$ [MeV] & Ghost config.& Gluon config. & Plaq. config. & 1-$<P>/P_\infty$ & p & Sweeps\cr
\hline
2.02 & 20 & 30 & 610 & 1206 & 1874 & 2430 & 11525 & 369211 & -5(4) 10$^{-6}$ & 0.83 & 30 \cr 
\hline
3.56 & 20 & 10 & 347 & 685 & 1064 & 2102 & 12319 & 355257 & 1(1) 10$^{-5}$ & 0.84 & 30 \cr
\hline
4.04 & 40 & 30 & 306 & 610 & 961 & 1964 & 10579 & 527689 & 1(2) 10$^{-6}$ & 0.88 & 50 \cr
\hline
6.06 & 60 & 30 & 204 & 408 & 644 & 1723 & 7311 & 510688 & 0(1) 10$^{-6}$ & 0.93 & 70 \cr
\hline
7.11 & 40 & 10 & 174 & 347 & 546 & 2161 & 10758 & 536786 & -2(6) 10$^{-6}$ & 0.87 & 50 \cr
\hline
8.08 & 80 & 30 & 153 & 306 & 484 & 1429 & 4898 & 438579 & 0(1) 10$^{-6}$ & 0.90 & 90 \cr
\hline
10.1 & 100 & 30 & 123 & 245 & 387 & 747 & 1988 & 216391 & -3(1) 10$^{-6}$ & 0.96 & 110 \cr
\hline
10.7 & 60 & 10 & 116 & 232 & 366 & 1825 & 7108 & 496291 & -2(4) 10$^{-6}$ & 0.92 & 70 \cr
\hline
12.1 & 120 & 30 & 102 & 204 & 323 & 552 & 1754 & 225036 & 1(1) 10$^{-6}$ & 0.95 & 130 \cr
\hline
14.1 & 140 & 30 & 87.6 & 175 & 371 & 368 & 368 & 53971 & -2(2) 10$^{-6}$ & 0.97 & 150 \cr 
\hline
14.2 & 80 & 10 & 87.0 & 174 & 275 & 1582 & 6465 & 579900 & -1(3) 10$^{-6}$ & 0.92 & 90 \cr
\hline
16.2 & 160 & 30 & 76.6 & 153 & 325 & 291 & 291 & 48199 & 0(2) 10$^{-6}$ & 0.98 & 170 \cr 
\hline
17.8 & 100 & 10 & 69.6 & 139 & 220 & 1339 & 4337 & 478853 & -2(3) 10$^{-6}$ & 0.96 & 110 \cr
\hline
18.2 & 180 & 30 & 68.1 & 136 & 289 & 308 & 308 & 56724 & -2(1) 10$^{-6}$ & 0.96 & 190 \cr
\hline
20.2 & 200 & 30 & 61.3 & 123 & 260 & 199 & 199 & 40584 & -2(1) 10$^{-6}$ & 0.96 & 210 \cr
\hline
21.3 & 120 & 10 & 58.0 & 116 & 183 & 762 & 5236 & 678065 & 1(2) 10$^{-6}$ & 0.93 & 130 \cr
\hline
22.2 & 220 & 30 & 55.7 & 111 & 236 & 232 & 232 & 51577 & 1(1) 10$^{-6}$ & 0.99 & 230 \cr
\hline
24.2 & 240 & 30 & 51.1 & 102 & 217 & 232 & 232 & 55691 & 0(1) 10${-6}$ & 0.98 & 250 \cr
\hline
24.9 & 140 & 10 & 49.7 & 99.4 & 211 & 517 & 517 & 76053 & 1(5) 10$^{-6}$ & 0.96 & 150 \cr
\hline
28.4 & 160 & 10 & 43.5 & 87.0 & 184 & 455 & 455 & 75500 & -8(4) 10$^{-6}$ & 0.97 & 170 \cr
\hline
32.0 & 180 & 10 & 38.7 & 77.3 & 164 & 390 & 390 & 72034 & -4(4) 10$^{-6}$ & 0.97 & 190 \cr
\hline
35.6 & 200 & 10 & 34.8 & 69.6 & 148 & 328 & 328 & 66976 & -4(4) 10$^{-6}$ & 0.97 & 210 \cr
\hline
39.1 & 220 & 10 & 31.6 & 63.3 & 134 & 287 & 287 & 63703 & 3(3) 10$^{-3}$ & 0.98 & 230 \cr
\hline
42.7 & 240 & 10 & 29.0 & 58.0 & 123 & 394 & 394 & 96075 & 0(2) 10$^{-6}$ & 0.98 & 250 \cr 
\hline
\end{tabular}
\end{ruledtabular}
\end{table*}

The generation of configurations in two dimensions and their gauge-fixing to Landau gauge can be and has been done exactly as in higher dimensions \cite{Cucchieri:2006tf}. In particular, the confirmation of the Gribov-Zwanziger scenario in the present work implies that the problem of Gribov-Singer copies \cite{Gribov,Singer:dk} should also in two dimensions become irrelevant for Green's functions in the infinite volume limit \cite{Zwanziger:2003cf}: Gribov-Singer effects should become smaller with increasing volume. Hence they have been ignored here, although, as discussed in section \ref{sprop}, effects at finite volume cannot be excluded.

To give units to the momenta, the infinite volume limit of the string tension for a given $\beta$, which can be determined analytically \cite{Dosch:1978jt}, is set to (440 MeV)$^2$. The configurations used are shown in table \ref{conf}. The comparison with the (also exactly known) infinite volume value of the plaquette \cite{Dosch:1978jt} shows that locally the continuum has been reached. However, the discussion in section \ref{sprop} shows that this is not correct globally.

\section{Lattice artifacts other than finite volume}\label{aarti}

\begin{figure*}
\includegraphics[width=0.5\linewidth]{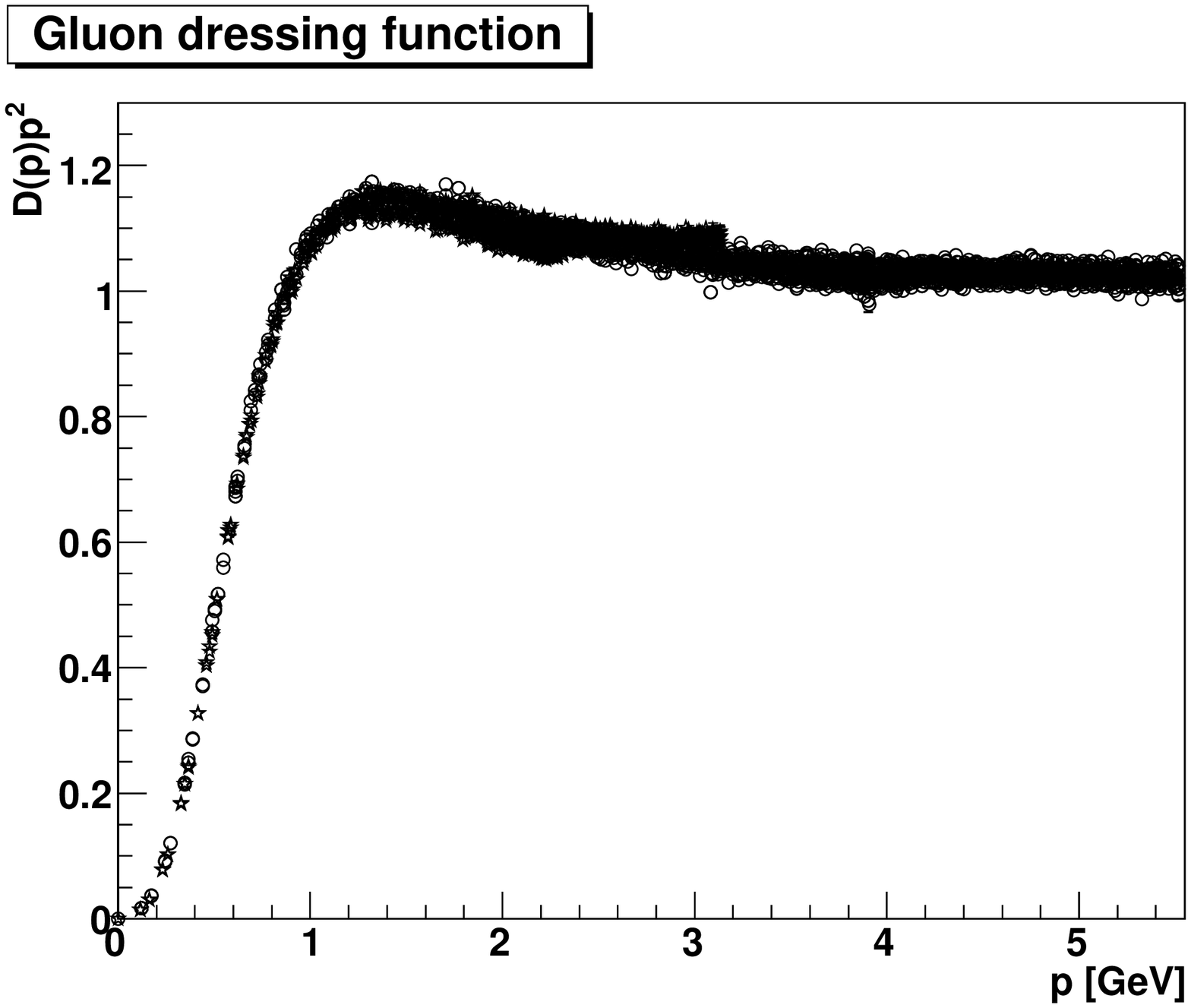}\includegraphics[width=0.5\linewidth]{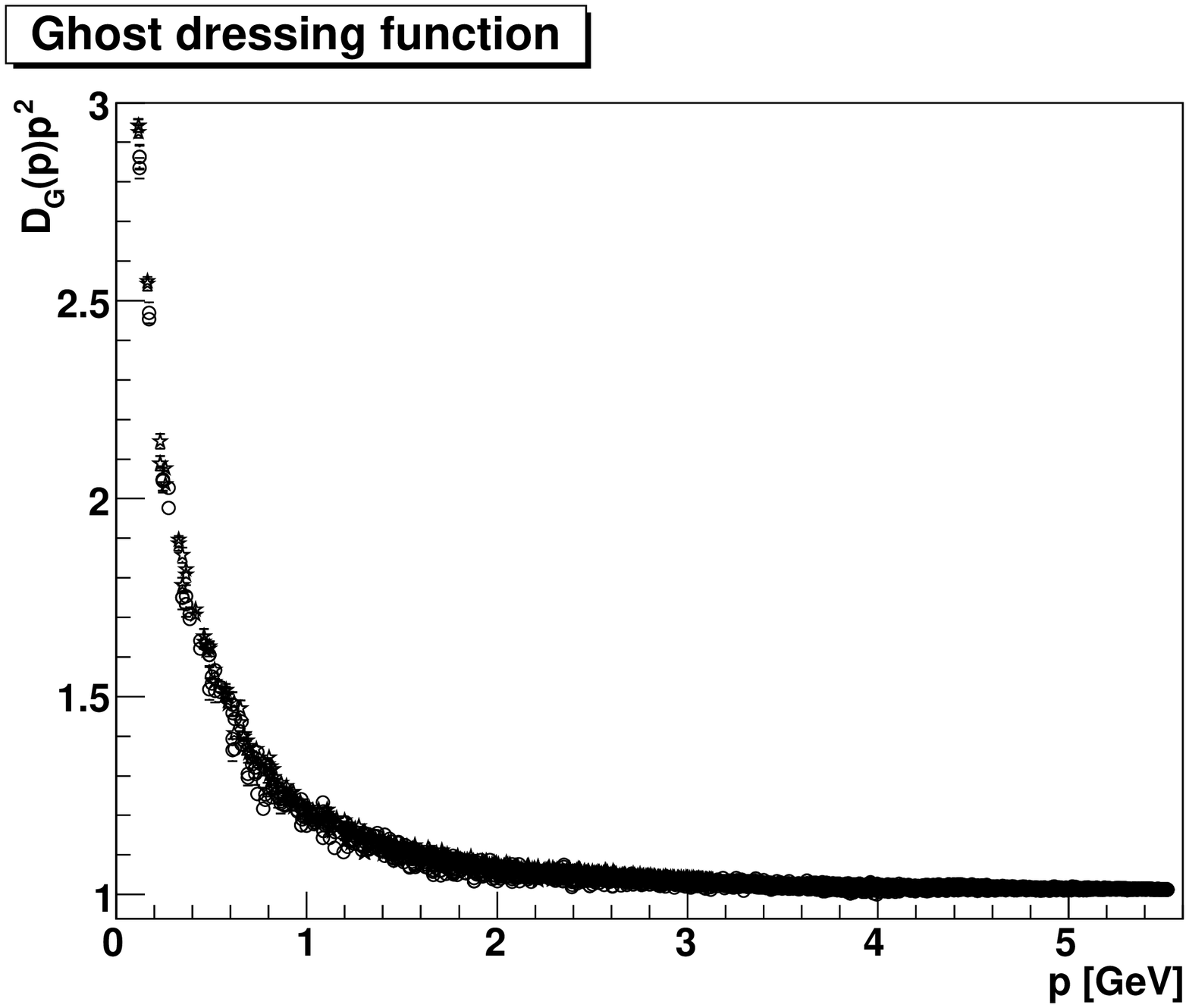}
\caption{\label{fdis}Consequences of different discretizations and violation of rotational invariance in case of the gluon (left panel) and ghost (right panel) dressing functions. Open circles correspond to a system at $\beta=30$ and a volume of (10.1 fm)$^2$, open stars to a system at $\beta=10$ and a volume of (10.7 fm)$^2$. The different momentum directions have not been marked differently.}
\end{figure*}

As one of the main claims here is that the deviation from the asymptotic continuum form in the infrared is a pure finite-volume effect, it is necessary to check the influence of other lattice artifacts. In particular, discretization effects and violation of rotational symmetry may be relevant. The latter is known to be a significant effect when comparing correlation functions measured along different directions of the hypercube (see, e.\ g., \cite{Cucchieri:2006xi}), in the present case along an edge or along a diagonal. In figure \ref{fdis}, these effects are explicitly checked. The results are at roughly the same volume of about (10.3 fm)$^2$ at two different $\beta$s, 10 and 30, and results with momenta along any possible direction are directly compared.

It is clearly visible that, despite a factor of nearly 2 in $a$, both results agree remarkably well over the whole range of momenta. Thus discretization effects are nearly negligible, at least for a volume of a few fm$^2$ and momentum not too close to the maximum one. Treating only the physical volume as an independent parameter in the infrared throughout the main text is hence justified. Also no significant effect is seen of the violation of rotational invariance, which is usually most pronounced at intermediate momenta in the gluon dressing function. For the current case a few tens of lattice sites along each edge seem to be sufficient to have already a quite good approximation of rotational invariance.

Furthermore, there is no distinct difference between the gluon and the ghost dressing function in terms of these artifacts. In case of the propagators these effects would be even diminished, as the trivial factor $p^{-2}$ helps in the reduction of such artifacts. Hence the totally dominant contribution for the artifacts in the correlation functions in the infrared is clearly the finite physical volume.

Similar observations pertain to all quantities measured here, and hence only the physical volumes are used as explicit parameters in the main text, and no heed is paid for the different $\beta$-values. The only exception observed here is in the case of the running coupling in section \ref{src}, where an overall scaling factor has been seen. This issue has been discussed in detail in this section \ref{src}.

\section{Contributions in other color tensor structures}\label{atensor}

\begin{figure*}
\includegraphics[width=0.5\linewidth]{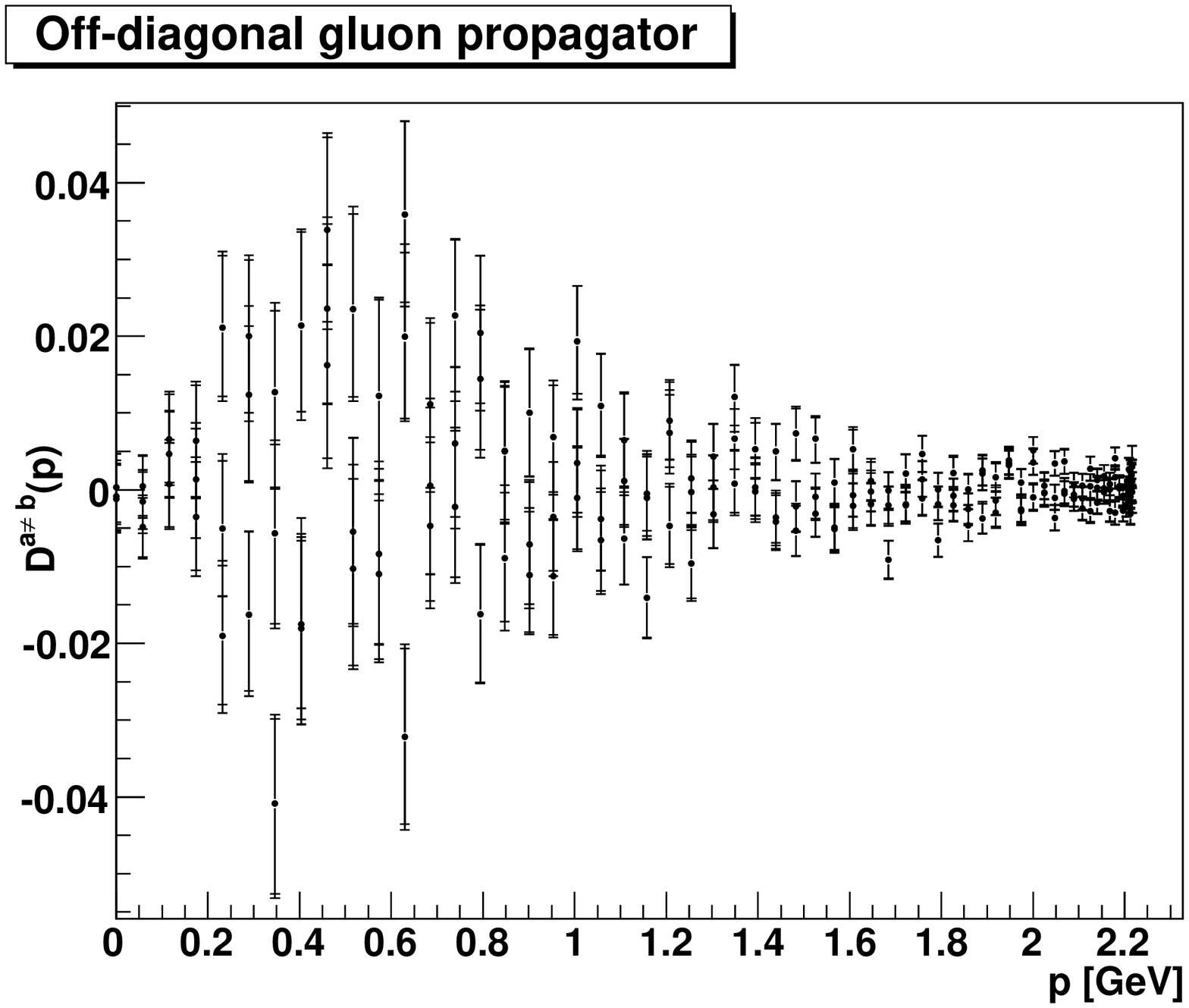}\includegraphics[width=0.5\linewidth]{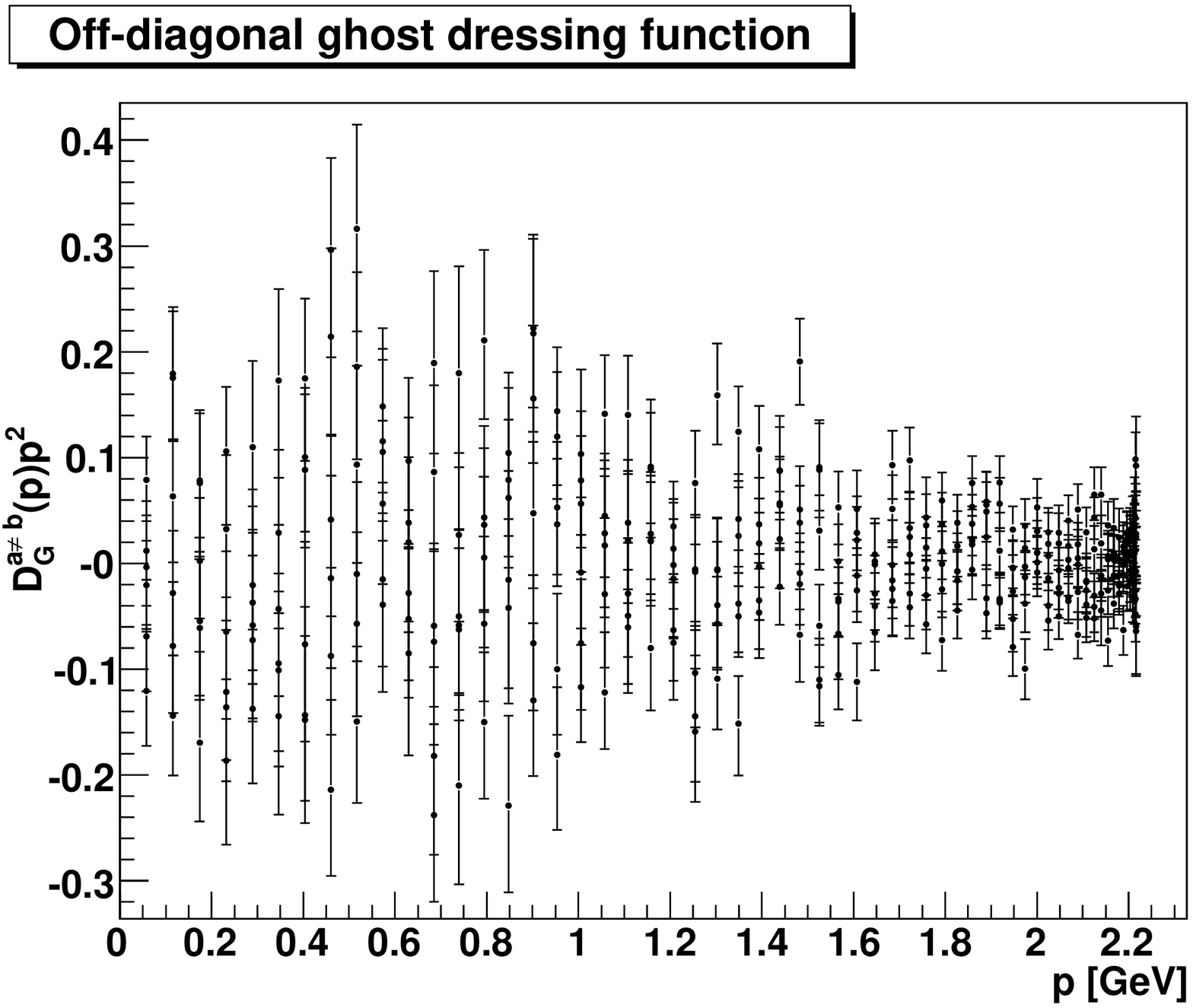}
\caption{\label{foff}The color off-diagonal elements of the gluon propagator (left) and of the ghost propagator (right) on a (21.3 fm)$^2$ volume.}
\end{figure*}

There is no a-priori necessity for correlation functions to carry only their tree-level color structure, although such a color structure permits a consistent solution using functional methods in the infrared, at least in four dimensions \cite{Fischer:2007pf}. Therefore, this property should be explicitly checked. This is done for the ghost and the gluon propagator in figure \ref{foff}. All contributions are compatible with zero. Furthermore, the average value decreases in all cases with increasing statistics. So, within the statistics available, there are no color-off-diagonal components in the propagators. Due to the structure of the DSEs, it is then very hard to imagine how the higher $n$-point Green's functions should have a color structure different from the tree-level one. This can, of course, not be excluded by this result. Nonetheless, it seems to be unlikely.


\end{document}